\documentclass[preprint,aps,amsmath,amssymb,nofootinbib,showpacs,
tightenlines]{revtex4}
\usepackage{graphicx}
\newcommand{\ba}{\begin{eqnarray}}
\newcommand{\ea}{\end{eqnarray}}
\newcommand{\nn}{\nonumber}

\newcommand{\occs}
{\begin{picture}(10,8)
\put(4,2){\circle*{4}}
\end{picture}}

\newcommand{\cons}
{\begin{picture}(11,8)
\put(5,3){\circle*{4}}
\thicklines
\put(4,-1){\line(1,4){2.5}}
\end{picture}}

\newcommand{\pina}{\begin{picture}(10,10)
\multiput(2,4)(6,0){2}{\circle*{4}}
\end{picture}}

\newcommand{\pinb}{\begin{picture}(15,10)
\multiput(2,4)(6,0){3}{\circle*{4}}
\end{picture}}

\newcommand{\pinc}{
\begin{picture}(10,15)
\multiput(2,9)(6,0){2}{\circle*{4}}
\put(8,1){\circle*{4}} 
\end{picture}}

\newcommand{\pind}{
\begin{picture}(10,15)
\multiput(2,9)(6,0){2}{\circle*{4}}
\multiput(2,1)(6,0){2}{\circle*{4}} 
\end{picture}}

\newcommand{\pine}{
\begin{picture}(15,15)
\multiput(2,9)(6,0){3}{\circle*{4}}
\put(8,1){\circle*{4}}
\end{picture}}

\newcommand{\pinf}{
\begin{picture}(15,15)
\multiput(2,9)(6,0){3}{\circle*{4}}
\put(14,1){\circle*{4}}
\end{picture}}

\newcommand{\ping}{
\begin{picture}(15,15)
\multiput(2,9)(6,0){2}{\circle*{4}}
\multiput(8,1)(6,0){2}{\circle*{4}} 
\end{picture}}

\newcommand{\pini}{
\begin{picture}(15,15)
\multiput(2,9)(6,0){2}{\circle*{4}}
\put(14,1){\circle*{4}} 
\end{picture}}

\newcommand{\pinj}{
\begin{picture}(15,15)
\put(2,9){\circle*{4}}
\put(8,1){\circle*{4}} 
\end{picture}}

\newcommand{\inda}{\begin{picture}(10,10)
\put(2,4){\circle*{4}}
\thicklines
\put(8,4){\circle*{4}}
\put(7,0){\line(1,4){2.5}}
\end{picture}}

\newcommand{\indb}{\begin{picture}(15,10)
\multiput(2,4)(6,0){3}{\circle*{4}}
\thicklines
\multiput(7,0)(6,0){2}{\line(1,4){2.5}}
\end{picture}}

\newcommand{\indbuns}{\begin{picture}(15,10)
\multiput(2,4)(12,0){2}{\circle*{4}} 
\put(6,2){-}
\thicklines
\put(13,0){\line(1,4){2.5}}
\end{picture}} 

\newcommand{\indc}{
\begin{picture}(10,15)
\multiput(2,9)(6,0){2}{\circle*{4}}
\thicklines
\put(7,5){\line(1,4){2.5}}
\put(8,1){\circle*{4}} 
\put(7,-3){\line(1,4){2.5}}
\end{picture}}

\newcommand{\indd}{
\begin{picture}(10,15)
\multiput(2,9)(6,0){2}{\circle*{4}}
\thicklines
\put(7,5){\line(1,4){2.5}}
\multiput(2,1)(6,0){2}{\circle*{4}} 
\multiput(1,-3)(6,0){2}{\line(1,4){2.5}}
\end{picture}}

\newcommand{\inde}{
\begin{picture}(15,15)
\thicklines 
\multiput(2,9)(6,0){3}{\circle*{4}}
\multiput(7,5)(6,0){2}{\line(1,4){2.5}}
\put(8,1){\circle*{4}}
\put(7,-3){\line(1,4){2.5}}
\end{picture}}

\newcommand{\indf}{
\begin{picture}(15,15)
\thicklines 
\multiput(2,9)(6,0){3}{\circle*{4}}
\multiput(7,5)(6,0){2}{\line(1,4){2.5}}
\put(14,1){\circle*{4}}
\put(13,-3){\line(1,4){2.5}}
\end{picture}}

\newcommand{\indg}{
\begin{picture}(15,15)
\multiput(2,9)(6,0){2}{\circle*{4}}
\thicklines
\put(7,5){\line(1,4){2.5}}
\multiput(8,1)(6,0){2}{\circle*{4}} 
\multiput(7,-3)(6,0){2}{\line(1,4){2.5}}
\end{picture}}

\newcommand{\indh}{\begin{picture}(22,10)
\multiput(2,4)(6,0){4}{\circle*{4}}
\thicklines
\multiput(7,0)(6,0){3}{\line(1,4){2.5}}
\end{picture}}

\newcommand{\indhuns}{\begin{picture}(22,10)
\multiput(2,4)(18,0){2}{\circle*{4}}
\put(6,2){- -}
\thicklines
\put(19,0){\line(1,4){2.5}}
\end{picture}}

\newcommand{\indi}{
\begin{picture}(15,15)
\multiput(2,9)(6,0){2}{\circle*{4}}
\thicklines
\put(7,5){\line(1,4){2.5}}
\put(14,1){\circle*{4}} 
\put(13,-3){\line(1,4){2.5}}
\end{picture}}

\newcommand{\indk}{
\begin{picture}(15,15)
\multiput(2,9)(6,0){2}{\circle*{4}}
\thicklines
\put(7,5){\line(1,4){2.5}}
\put(2,1){\circle*{4}} 
\put(1,-3){\line(1,4){2.5}}
\end{picture}}

\newcommand{\inddot}{\begin{picture}(25,20)
\multiput(2,14)(6,0){2}{\circle*{4}}
\thicklines
\put(7,10){\line(1,4){2.5}}
\multiput(12,14)(3,0){3}{\circle*{1}}
\put(22,14){\circle*{4}}
\put(21,10){\line(1,4){2.5}}
\put(14.5,11){\makebox(0,0)[t]{$\underbrace{}$}}
\put(16,-2){\makebox(0,0)[b]{\tiny k-times}}
\end{picture}}

\begin{document}

\title{On-lattice coalescence and annihilation of immobile reactants 
in loopless lattices and beyond}
\author{E. Abad} 
\email{eabad@ulb.ac.be}
\affiliation{Centre for Nonlinear Phenomena and Complex Systems, 
Universit\'e Libre de Bruxelles CP 231, 1050 Bruxelles, Belgium.}

\begin{abstract}
We study the behavior of the  
chemical reactions $A+A\to A+S$ and $A+A\to S+S$ (where the reactive
species $A$ and the inert species $S$
are both assumed to be immobile) embedded on 
Bethe lattices of arbitrary coordination number $z$ and on a two-dimensional
($2D$) square 
lattice. For the Bethe lattice case, exact solutions for the coverage in
the $A$ species in terms of the initial condition
are obtained. In particular, our results hold for the 
important case of an infinite one-dimensional ($1D$) lattice ($z=2$).
The method is based on an expansion in terms of conditional probabilities
which exploits a Markovian property of these systems. 
Along the same lines, an approximate solution for the case of a 
$2D$ square lattice is developed. The effect of dilution in a random
initial condition is discussed in detail, both for the lattice 
coverage and for the spatial distribution of reactants. 
\end{abstract}

\maketitle

\section{Introduction}

A rigorous description of
the dynamics of the relevant macrovariables in reaction-diffusion 
systems requires a probabilistic multilevel approach 
retaining the essential features of the underlying many-body problem
\cite{vkamp,gard,nicpr}. In this coarse-grained picture, typical 
macrovariables such 
as concentrations are no longer deterministic, but rather stochastic 
quantities. In a number of typical situations, the equations governing 
the dynamics of the mean concentrations turn out to be identical with the 
classical law of mass action. In the absence of external asymmetries or 
of symmetry-breaking instabilities, the latter can be regarded as a 
mean-field (MF) law, in the sense that each part of the system is assumed 
to interact 
with the whole bulk at all times by means of an effective field which 
does not account for spatial effects.

The above classical approach can be 
regarded as a good approximation as long as the characteristic time
associated with the mean free path is short compared to the mean reaction
time within the typical interaction radius. This is only the case if the 
system is well mixed at all times, either through external stirring or 
through fast internal diffusion. The opposite situation corresponds to 
the diffusion-controlled limit, where each reactant typically explores a 
significant portion of space before undergoing a reactive collision, and 
the way in which reactants are 
distributed on a microscopic scale starts to become important
for the determination of macrovariables such as global concentrations or, 
in the case of a lattice systems, the coverage in the different species.    
In such cases, classical MF approaches fail to describe 
the onset of inhomogeneous fluctuations induced by the intrinsic 
chemical noise of the system. Such fluctuations are nowadays
directly observable at nanometric scales with the help of STM and FIM 
microscopy techniques \cite{such0,such1} and can be enhanced by specific 
geometric constraints 
and/or in low dimensions (e.g. catalytic surfaces), where 
external stirring is difficult and diffusion inefficient; eventually,
they may give rise to non-classical effects such as memory of the 
initial condition, self-ordering phenomena, etc. 
\cite{havlin}. Elucidating the 
role of geometry in this context is of great theoretical and practical
interest in view of the recent progress in the development of 
nanoscale supports. 

Fluctuation-induced effects become even stronger in systems 
with immobile reactants, the object of the present paper. The particular
systems we shall investigate here are the on-lattice 
reactions $A+A \rightarrow A+S$ 
and $A+A \rightarrow S+S$ 
with nearest neighbor interactions, where $A$ 
and $S$ denote, respectively, a site occupied by the reactive species
(``occupied site'') and the inert species (``empty site''), both assumed
to be immobile. Popularly, 
these reactions are termed coalescence (CR) and annihilation reaction (AR) 
respectively. Various workers have intensively investigated the
CR \cite{avrabursch,Spou,abad2} and the AR \cite{Spou,Torn,Lus,Schut,Gryn,Fam2}
in the diffusion-controlled limit. Besides a series of applications for 
nucleation and aggregation systems \cite{Leyv,prov2}, the diffusion-controlled 
CR has also been recently used as a model for exciton fusion in 
polymers and molecular crystals \cite{kop0,kop1,privb}, 
while the AR model provides a basic description for recombination 
processes and exciton annihilation \cite{Racz,privb}.
In the immobile reactant limit, the AR model 
has been e.g. used to study free radical recombination 
on surfaces \cite{Jack}, cyclization reactions in polymers \cite{cohen} 
and colloid deposition problems \cite{Bart} among other applications. 
Note the formal 
similarity of this model and models for dimer 
random sequential absorption (DRSA) \cite{Evans,Dick}. In such DRSA models, 
the deposition of a dimer on two empty sites is dual to the 
removal of two neighboring particles from the lattice upon reactions
in the AR model, i.e. empty sites play the role of occupied sites and 
vice versa. 
There exists also a (less obvious) mapping between the CR model and
a particular case of random monomer filling with nearest neighbor
cooperativity \cite{kell,mity,fan}. However, most studies concerning the above 
RSA models were performed for a fixed initial condition.
Typically, the latter corresponds to a situation where all
lattice sites are vacant, which in the dual picture of our model
is equivalent to a fully covered lattice. 
In contrast, we shall consider here the general case in 
which the lattice is partially filled initially and study how this 
affects the subsequent dynamics and steady state of the system. 

Previous studies have shown that in the immobile reactant limit the 
one-dimensional CR and AR models 
with nearest neighbor interactions are characterized by an exponential 
decay of the mean coverage $\theta(t)$ in the reactive species to a nonergodic 
set of invariant states, as opposed to the empty state predicted by the 
MF equation \cite{baras,abad1}. In the present work, we extend these results to the 
case of a partially filled Bethe lattice with arbitrary coordination number. 
In such loopless lattices, the relevant hierarchy of probabilities can be 
truncated exactly using a shielding (Markovian) property of the conditional 
probabilities for the state of a given site. This method is used to  
generalize previous results
by Evans \cite{Evans3} and by Majumdar and Privman \cite{Priv}. 
Next, we treat the case of a $2D$ square lattice by  
performing an expansion based on the shortest unshielded path 
approximation developed by Nord and Evans for a series of RSA models 
\cite{Nord}. The results for the asymptotic coverage 
are then compared with Monte Carlo (MC) 
simulations. The effect of the initial condition on the spatial ordering 
induced by the reaction is also discussed by studying
the dynamical behavior of the conditional probabilities 
and the associated fluctuation correlations. 
In the last section, we summarize the main conclusions 
and outline possible extensions of our work.   

\section{The CR and the AR model in Bethe lattices: 
exact solution via shielding property}

As a starting point, we consider an ensemble of Bethe lattices with
coordination number $z$ (the case $z=2$ corresponds to an infinite 
$1D$ lattice). In each lattice, sites are initially occupied at random with 
probability $p$ (equal to the initial lattice coverage $\theta(0)$). 
We then let the particles interact according to the CR (AR) scheme 
with nearest neighbor interactions specified above.
By construction, the resulting statistical system will be translationally 
invariant at all times. Let us absorb the 
reaction rate $R$ into the time scale by introducing the dimensionless
time variable $\tau\equiv R\,t$. Let us denote by $P_k(\tau)$ the probability 
that $k$ randomly chosen nearest neighbor sites in a given 
lattice are all simultaneously occupied ($k$-site cluster). 
The evolution equations for the ensemble probabilities $P_k$ read \cite{Priv} 

\begin{equation}
\label{bethepriv}
\frac{dP_k}{d\tau}=-(k-1)P_k-\frac{\nu}{2}(zk-2k+2)\, 
P_{k+1},\quad k=1,2,\ldots
\end{equation}
where $\nu=1,2$ for the CR and the AR model, respectively.
 
The first term on the right hand side represents the destruction 
of a $k$-site cluster by interaction between two particles inside the 
cluster and is proportional to the number of internal bonds ($=k-1$ bonds). 
The second term represents the destruction of a $k$-site cluster due to 
the disappearance of a particle inside the cluster upon 
interaction with a neighboring particle just outside the cluster. 
Such an event is only possible if a $k+1$-site cluster preexists, 
implying that this term is proportional to $P_{k+1}$. Its coefficient
is proportional to the number of bonds between
the $k$ cluster sites and external neighboring sites (=$zk-2k+2$). 
The different value of 
$\nu$ stems from the fact that the $k$-site cluster is only destroyed 
if the particle vanishes upon interaction with a filled neighbor 
outside the cluster. In the CR case, this only happens one time 
out of two, since in a single event the particle inside the cluster has 
the same probability of vanishing as the neighbor particle outside the 
cluster.

We now seek a special solution of the hierarchy (\ref{bethepriv}) corresponding 
to our initial condition. One easily check that in our case $P_k(0)=p^k$. 
As it turns out, the hierarchy can be exactly truncated after the first two equations 

\begin{subequations}
\label{2eq}
\ba
\label{2eqset}
\frac{dP_1}{d\tau}&=&-\frac{\nu}{2} \, z P_2, \\
\label{2eqset2}
\frac{dP_2}{d\tau}&=&-P_2-\nu\, (z-1) P_3.
\ea
\end{subequations}
Exact truncation is possible because of the existence of a special property for  
the quantity $Q_{\inddot}(\tau)=P_{k+1}/P_k $, i.e. 
the conditional probability that a site is occupied
given that its $k$ nearest neighbors along any irreversible path starting from 
the site are occupied. Here ``\occs'' denotes an occupied site, 
while ``\cons'' denotes a 
conditioning occupied site. Rewriting Eqs. (\ref{2eq}) in terms of these 
probabilities, one has

\begin{subequations}
\label{condhie}
\ba
\frac{d}{d\tau}\ln Q_{\occs}&=&-\,\nu\, \frac{z}{2} \, Q_{\inda}, \\
\frac{d}{d\tau}\ln Q_{\inda}&=&-1+\frac{\nu z}{2}Q_{\inda} - \nu\,(z-1) 
Q_{\indb},   
\ea
\end{subequations}

where the notation $Q_{\occs}\equiv P_1$ has been used. 
Now, in a Bethe lattice a conditioning site specified as occupied ``shields'', i.e. 
clusters belonging to disjoint irreversible paths starting from the occupied
site evolve independently of each other \footnote{ 
In the special case of a $1D$ lattice ($z=2$), this means that sites to the left of
the occupied site do not ``see'' those on the right hand side.}.
As a consequence, one has \cite{Vette,mity,Evans3}

\begin{equation}
\label{prop1}
Q_{\inda}=Q_{\indb}=Q_{\indh}\cdots,\quad \mbox{etc.}
\end{equation}
 
i.e. the memory of the system is limited to the nearest neighbor site. 
In this sense, the subset of occupied sites can be said to display 
(first-order) spatial markovianity \cite{Malek}. 

Using the shielding property (\ref{prop1}), Eqs. (\ref{condhie}) 
become a closed 
two-variable system. Since the system is translationally invariant, the 
local probability $Q_{\occs}$ is identical with the global coverage $\theta$.  
The solution of Eqs. (\ref{condhie}) reads

\begin{subequations}
\label{condbeth}
\ba
\label{betheta}
Q_{\occs}(\tau)&=&\theta(\tau)=p\,\left[1+\frac{(z-2)}{2}\,\nu\,p
(1-e^{-\tau})\right]^{-z/(z-2)}, \\
Q_{\inda}(\tau)&=&\frac{p\, e^{-\tau}}
{1+\frac{\textstyle (z-2)}{\textstyle 2} 
\,\nu\,p\,(1-e^{-\tau})}.
\ea
\end{subequations}

Thus, the global coverage attains the asymptotic value

\begin{equation}
\theta_S=\theta(\infty)=p\,\left[1+\frac{z-2}{2}\,\nu\,p\right]^{-z/(z-2)}.
\end{equation}

Regardless of the value of $z$, $\theta_S$ increases
monotonically with increasing $p$ for the CR model, while in the AR 
model it follows a non-monotonic behavior with a universal 
maximum at $p=\frac{1}{2}$. This may be related to the fact 
that the mean asymptotic number 
of particles yielded by islands (=disconnected clusters)
created by the ongoing reactions does not grow monotonically with 
the island size, as opposed to the CR case \cite{prov3,thes}. Thus, larger 
islands characteristic of high values of the initial coverage $p$ 
may eventually yield a lower number of particles than smaller ones. 
In particular two-particle islands are known to disappear from the 
system, while one-particle islands survive forever. 

As expected, $\theta_S$ decreases 
strongly as a function of $z$, approaching 
a zero value when $z\to\infty$. This is in agreement with our intuitive 
expectation that the system must approach the classical MF prediction 
with increasing connectivity. 

Eq. (\ref{betheta}) generalizes previous results by Evans 
for the DRSA problem equivalent to the $p=1$ case \cite{Evans3}
and Majumdar and Privman for the $\nu=2$ case \cite{Priv}.
The special case of a $1D$ lattice is obtained when $z\to 2^+$.
In this limit, Eqs. (\ref{condbeth}) become

\begin{subequations}
\ba
\label{expcov}
\theta(\tau)&=&p\, \exp{(\,\nu\,p\,[e^{-\tau}-1]\,)}, \\
Q_{\inda}(\tau)&=&p\, e^{-\tau}.
\ea
\end{subequations}

The conditional probability $Q_{\inda}$ 
is the same for both reaction schemes, as opposed to the 
$\nu$-dependent lattice coverage. The latter approaches the 
nonvanishing asymptotic value  

\begin{equation}
\label{asres1D}
\theta_S\equiv \theta(\infty)=p\,e^{-\nu p}.
\end{equation}
in contrast to the prediction of the 
MF solution $\theta_{MF}(\tau)=p/(1+\nu\,p\,\tau)$. 
Note that for $\nu=2$ and $p=1$ the asymptotic coverage predicted by 
this equation is  
compatible with Flory's famous $e^{-2}$ prediction for the isomorphic 
dimer filling problem \cite{Flo2, Ken}.    

\section{Approximate expansion vs. ``exact`` Monte Carlo results on a $2D$ 
square lattice}

We now turn to the task of finding a suitable approximation scheme
for the AR and the CR in lattices containing loops. The complex topology 
of particle clusters does not allow for an exact solution in this case. 
One must therefore resort to truncated expansions in terms
of conditional probabilities and to MC simulations. 

As in the $1D$ case \cite{abad1}, 
an Ursell expansion of the cluster probabilities using 
fluctuation correlation functions does not yield good results here 
either, since multisite fluctuation correlations do 
not decay monotonically with
increasing number of sites and are therefore non-negligible (see 
subsection \ref{secor}). 
On the other hand, neglecting the cluster probabilities beyond a certain order
may provide good agreement with simulations in the low-$p$ regime, but
the agreement is much worse when $p\approx 1$. In order to obtain a reasonable
agreement in the whole range of $p$ values, we must refine the truncation
procedure inspired by a (spatial) Markovian property of the system 
analogous to the one observed for the $1D$ system: while in $1D$ 
a single occupied site disconnects the reactive dynamics in the left 
and right half lines, in $2D$ an infinite line of sites 
specified as occupied decomposes the lattice into two independently 
evolving sublattices. More generally, it can be shown that for two-site
processes such as the CR and the AR model in regular lattices of arbitrary 
Euclidean dimensionality ``hyperwalls'' of thickness equal to one lattice site
shield one side of the lattice from the other \cite{Evans}. 

Recently, Nord and Evans used this generalization of the 
Markovian property as a starting point to devise an expansion scheme 
in terms of conditional multisite probabilities for a DRSA model on a square 
lattice \cite{Nord}. The main idea 
is to neglect the influence of conditioning occupied sites beyond a certain
cut-off distance $d_c$ (measured in lattice spacings). However, the 
calculation of $d_c$ should be
tailored so as to reflect the shielding property of occupied sites. 
The effective distance between an occupied site and a \cons site should be
defined as the shortest ``unshielded'' path which is not blocked by 
other \cons sites. For instance, in the particular case to be studied here 
(an infinite $2D$ square lattice), the distance $d_c$ between the \occs site 
and the righmost \cons site associated with the (translationally invariant)
probabilities $Q_{\inda},Q_{\indb}$ and $Q_{\indg}$ is
respectively 1, 4 and 5 lattice spacings. 

The case studied by Nord and Evans corresponds to the AR model with 
an initially full lattice. We shall now extend their calculations to the 
AR and the CR models with an arbitrary initial lattice coverage $p$. 
The starting point to perform the expansion are again the evolution 
equations for clusters of occupied sites. The first few equations for the 
evolution of low-order clusters are 

\begin{subequations}
\label{clusteq2D}
\ba
\frac{dP_{\occs}}{d\tau}&=&-2\,\nu\, P_{\pina}, \\
\frac{dP_{\pina}}{d\tau}&=&-P_{\pina}-\nu\,P_{\pinb}-
2\nu\,P_{\pinc}, \\
\frac{dP_{\pinj}}{d\tau}&=&-2\nu\,P_{\pinc} -2\nu\,P_{\pini}, \\
\frac{dP_{\pinc}}{d\tau}&=&-2P_{\pinc}-\nu\,P_{\pinf}-
\nu\,P_{\pine}
-\nu\,P_{\ping}-\nu\,P_{\pind}, \\
\mbox{etc.}&& \nn
\ea
\end{subequations}
Rewriting the first four equations of this hierarchy in terms of the 
$Q$'s, we obtain

\begin{subequations}
\label{eqsQ}
\ba
\label{1stQ}
\frac{d}{d\tau}\ln Q_{\occs}&=&-2\,\nu\, Q_{\inda}, \\
\label{2ndQ}
\frac{d}{d\tau}\ln Q_{\inda}&=&-1-\nu\,Q_{\indb}-2\nu\,Q_{\indc}+2\nu\,
Q_{\inda}, \\
\frac{d}{d\tau}\ln Q_{\indc}&=&-1+\nu\,Q_{\indb}+2\nu\,Q_{\indc} \nn \\ 
\label{3rdQ}
&& -\nu\,Q_{\indf}-\nu\,Q_{\inde}-\nu\,Q_{\indg}-\nu\,
Q_{\indd}, \\ 
\frac{d}{d\tau}\ln Q_{\indk}&=&-2+2\nu\,Q_{\indk}+2\nu\,Q_{\indi} 
\nn \\ 
\label{4thQ}
&& -\nu\,Q_{\indf}-\nu\,Q_{\inde}-\nu\,Q_{\indg}-\nu\,Q_{\indd}. 
\ea
\end{subequations}

\subsection{First-order approximation}

Let us first consider the first-order approximation, i.e., we neglect those 
\cons sites beyond a distance further than one lattice spacing. We then 
have $Q_{\indb}\rightarrow Q_{\inda}, Q_{\indc}\rightarrow 
Q_{\inda}$, and  Eqs. (\ref{eqsQ}) lead to the closed
set of equations

\begin{subequations}
\ba
\frac{d}{d\tau}\ln Q_{\occs}&=&-2\,\nu\, Q_{\inda}, \\
\frac{d}{d\tau}\ln Q_{\inda}&=&-1-\nu\,Q_{\inda}.  
\ea
\end{subequations}
Taking into account the initial condition $Q_{\occs}(0)=Q_{\inda}(0)=p$,
these equations are readily integrated to obtain

\begin{subequations}
\ba
&& Q_{\occs}(\tau)=P_{\occs}(\tau)= 
\frac{p}{(1+\nu\,p\,(1-e^{-\tau}))^2}, \\
&& Q_{\inda}(\tau)=\frac{p\, e^{-\tau}}{1+\nu\,p\,(1-e^{-\tau})}.
\ea
\end{subequations}
Notice that this result for the 
lattice coverage $P_{\occs}$ and the conditional probability 
$Q_{\inda}$ is identical with the exact result 
in a Bethe lattice with $z=4$ (cf. Eq. (\ref{condbeth})). 

The 
asymptotic result

\begin{equation}
\label{asco2D}
\theta_S=P_{\occs}(\infty)=
\frac{p}{(1+\nu\,p)^2}
\end{equation}
can be expanded in powers of $p$ to obtain

\begin{equation}
\label{expascov}
\theta_S= p-2\nu\,p^2+3\nu^2p^3+{\cal O}(p^4).
\end{equation}
The different terms in the right hand side are recovered by a somewhat 
rougher truncation scheme
neglecting all cluster probabilities involving more 
than a given number of sites $k_{max}$ in equations (\ref{clusteq2D}). 
Note that the term in $p^2$ in the right hand side of the 
formula (\ref{expascov}) for the dilute case contains an additional 
factor $2$ with respect to the expansion of the $1D$ result (\ref{asres1D})
for small $p$. This suggest that, in a hypercubic lattice with coordination
number $z$, the prefactor of this term might have the form $\nu z/2$, as 
is the case in the corresponding expansion of Eq. (\ref{betheta}) for
a Bethe lattice.  

Let us now compare the asymptotic values of the coverage $\theta_S$
obtained from the first-order truncation of the cluster hierarchy 
with ``exact'' results from MC simulations.  
The MC algorithm for the AR and the CR model is performed as follows. 
At the beginning of each statistical 
realization the sites of a $N \times N$ periodic square lattice (torus) 
are randomly filled 
with particles until a predetermined global coverage $p$ is attained. 
The elementary time step $\delta t$ is chosen in such a a way that 
each lattice site is visited once on average after one time unit $\Delta t$, i.e.  
$\delta t\equiv \Delta t/N^2$. At each time step, a site $i$ and one of its four 
nearest neighbor sites $i'$ are chosen at random. If both are occupied, the reaction 
step takes place with probability $p_R=R\,\delta t=\delta \tau$ (where
$\delta \tau=\Delta \tau/N^2$), i.e. 
the particle at site $i$ is removed from the lattice in the CR case, while 
in the AR model both sites $i$ and
$i'$ are vacated. Fortunately, the convergence is rather fast with increasing 
linear size $N$ and number of statistical realizations $n_{real}$. 
The limiting values $\theta_S$ given in the tables 
\ref{TabCPD} and \ref{TabCTD} correspond to $N=200$ and 
$n_{real}=5000$ and its accuracy is at least equal to $10^{-5}$. 

\begin{figure}[ht]
\includegraphics[scale=0.5]{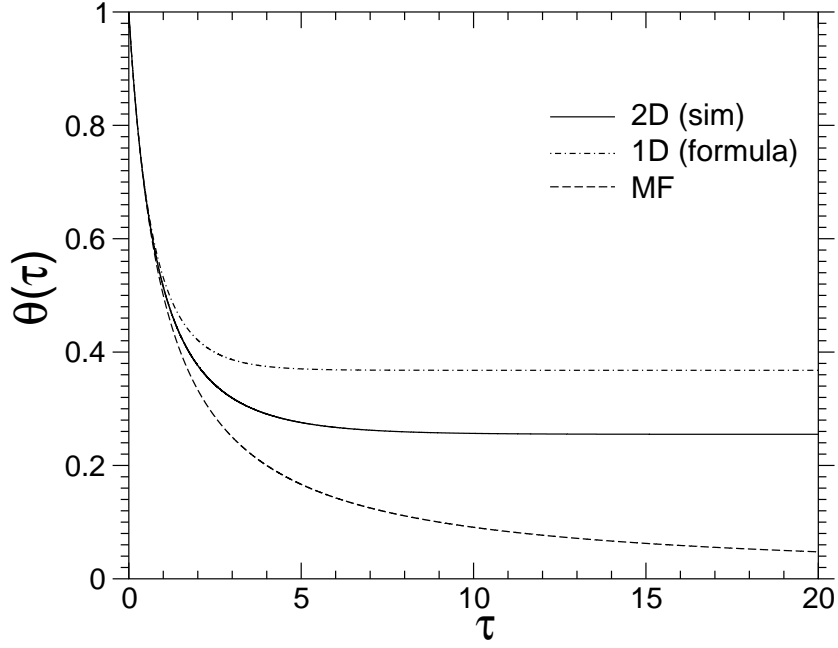}
\caption{\label{fig1b} Comparative plot displaying the analytical $1D$ 
solution for the coverage, the $2D$ simulation result on a square lattice
and the MF solution for the CR model.}
\end{figure}

Fig. \ref{fig1b} shows the coverage $\theta$ as a function of
the dimensionless time for the CR model. For an initially full lattice, 
a comparative
plot between the $1D$ solution, the MC result on the $2D$ square lattice 
and the simple MF approach is displayed. In the square lattice case, the mean 
coverage does not significantly change for times $\tau\approx 10$ and above, 
and its limiting value is found to be $0.2549$, 
about $30.7\%$ smaller than the $1D$ result
 $e^{-1}\approx 0.3679$. As expected, the higher connectivity of the 
$2D$ lattice ($z=4$, in contrast to $z=2$ for 
the $1D$ case) leads to an increased number of reactive events per 
occupied site, and the system gets closer to the empty state.
As in the $1D$ case, the long-time decay to the final state appears 
to be well fitted by an exponential.   

\begin{figure}[ht]
\includegraphics[scale=.5]{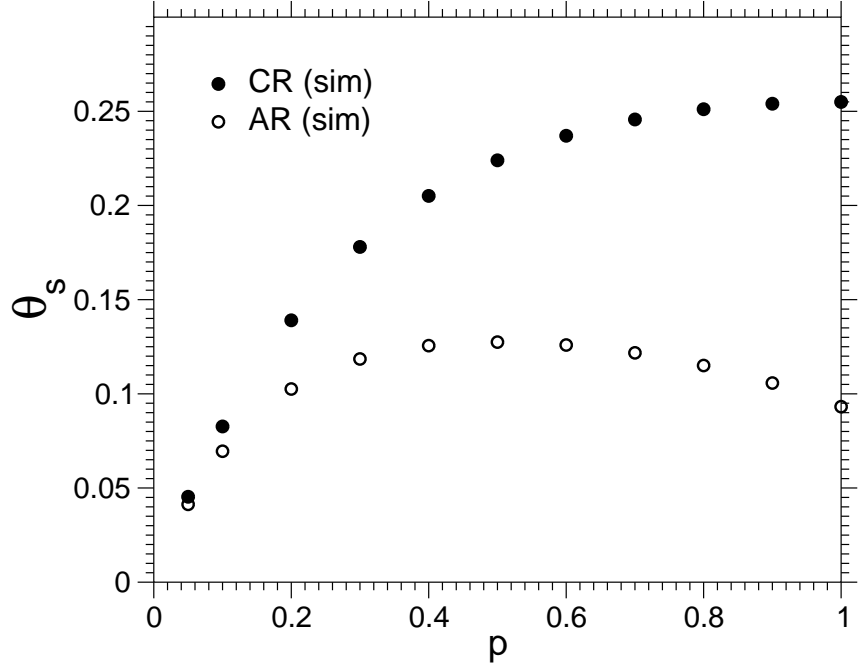}
\vspace*{0.5cm}
\caption{\label{fig2} Comparative plot showing the $p$-dependence of the 
asymtotic coverage for the CR and the AR models on a $2D$ square lattice.}
\end{figure}

\begin{figure}[ht]
\includegraphics[scale=0.5]{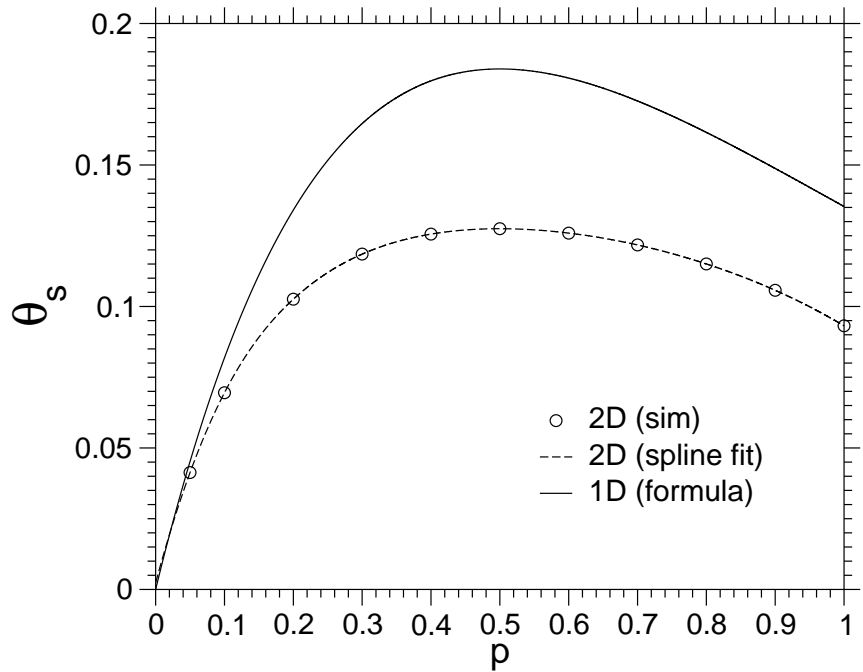}
\caption{\label{fig3} Nonmonotonic behavior of the final coverage
as a function of the initial coverage $p$ in the AR case. The continuous
line displays the $1D$ analytic result, while the dots correspond to 
simulation results on the $2D$ square lattice. The dot-dashed curve 
represents a spline fit of the MC results. }
\end{figure}

In the AR case, the simulation yields the exact value 
$\theta_S=0.0932$ for an initially full lattice, off by about 
$31.1\%$ from the exact value in $1D$. Fig. \ref{fig2} shows the 
stationary coverage 
$\theta_S$ as a function of $p$ for both the CR and the AR. 
The dependence is monotonic for the CR, whereas a maximum at $p=0.5$ 
is observed in the AR case. As in the case of
a Bethe lattice, this generic dependence on the initial coverage is 
likely to be robust in hypercubic lattices with arbitrary coordination
number (cf. Fig. \ref{fig3} for the AR case). 

\begin{figure}[ht]
\includegraphics[scale=0.5]{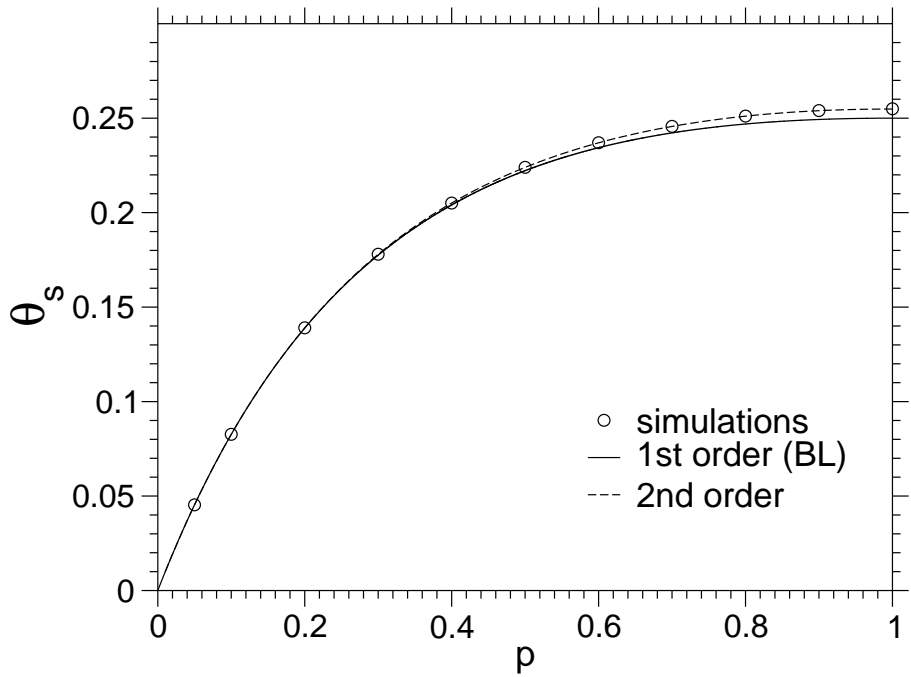}
\caption{\label{fig4} Exact simulation results for the CR vs. the first two
orders of the shortest unshielded path approximation (the first order 
corresponds to a Bethe lattice). Note the 
good agreement of the second order results with the simulation over
the whole $p$ range.}
\end{figure}

\begin{figure}[ht]
\includegraphics[scale=0.5]{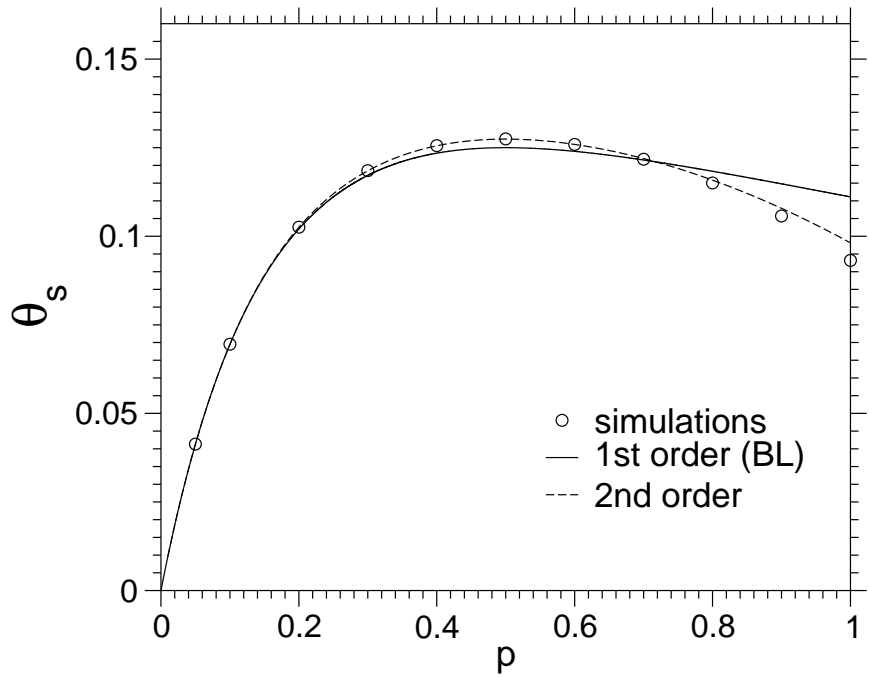}
\caption{\label{fig5} Exact simulation results for the AR vs. the first two
orders of the shortest unshielded path approximation (the first order 
corresponds to a Bethe lattice). Note that the 
agreement of the approximate results with the simulation in the 
saturation region $p\approx 1$ is worse than in the CR case. }
\end{figure}

For $p=1$ and the CR ($\nu=1$), we get 
$\theta_S=1/4=0.25$ from Eq. (\ref{asco2D}), which is smaller than 
the simulation value by $19.2\%$ (cf. Fig. \ref{fig4} and Table \ref{TabCPD}), 
whereas for $\nu=2$, 
the formula (\ref{asco2D}) yields $\theta_S=1/9=0.1111$, which is 
larger than the exact numerical value by $16.1\%$
(cf. Fig. \ref{fig5} and Table \ref{TabCTD}). Thus, for a sufficiently large
$p$, the first-order truncation (Bethe lattice solution) underestimates the 
asymptotic coverage in the CR case and overestimates it in the AR case. 

On the other hand, for sufficiently low 
values of $p$ the approximation gets better in both cases. 
Thus, for $p=0.5$ the simulation value is larger than the 
approximated one by just $0.8\%$ for the CR (cf. Table \ref{TabCPD}). 
The fact that, for a given order of truncation, the accuracy increases 
monotonically with $p$ in the parametric region $p \ll 1$ corresponding 
to a dilute system is by no means surprising: in the dilute limit
the $z=4$ Bethe lattice becomes a good approximation for the $2D$
square lattice, since ``lattice animals'' containing loops become rare.  

\subsection{Second-order approximation}
\label{secor}

For the second-order approximation, we take the whole set 
(\ref{eqsQ}) as a starting point and make the approximations

\[
Q_{\indb}, Q_{\indf}, 
Q_{\indi}\rightarrow Q_{\inda};\quad 
Q_{\inde},Q_{\indg} 
\rightarrow Q_{\indc};\quad Q_{\indd}\rightarrow Q_{\indk}, 
\]
thereby retaining all unshielded paths with lengths smaller than or equal 
to two lattice spacings. With this approximation we get from Eqs. 
(\ref{eqsQ})

\begin{subequations}
\label{secorhie}
\ba
\label{1stQ2}
\frac{d}{d\tau}\ln Q_{\occs}&=&-2\,\nu\, Q_{\inda}, \\
\label{2ndQ2}
\frac{d}{d\tau}\ln Q_{\inda}&=&-1+\nu\,Q_{\inda}-
2\nu\,Q_{\indc}, \\
\label{3rdQ2}
\frac{d}{d\tau}\ln Q_{\indc}&=&-1-\nu\,Q_{\indk}, \\
\label{4thQ2}
\frac{d}{d\tau}\ln Q_{\indk}&=&-2+\nu\,Q_{\inda}-
2\nu\,Q_{\indc}+\nu\,Q_{\indk}.
\ea
\end{subequations}

\begin{figure}[ht]
\includegraphics[scale=0.5]{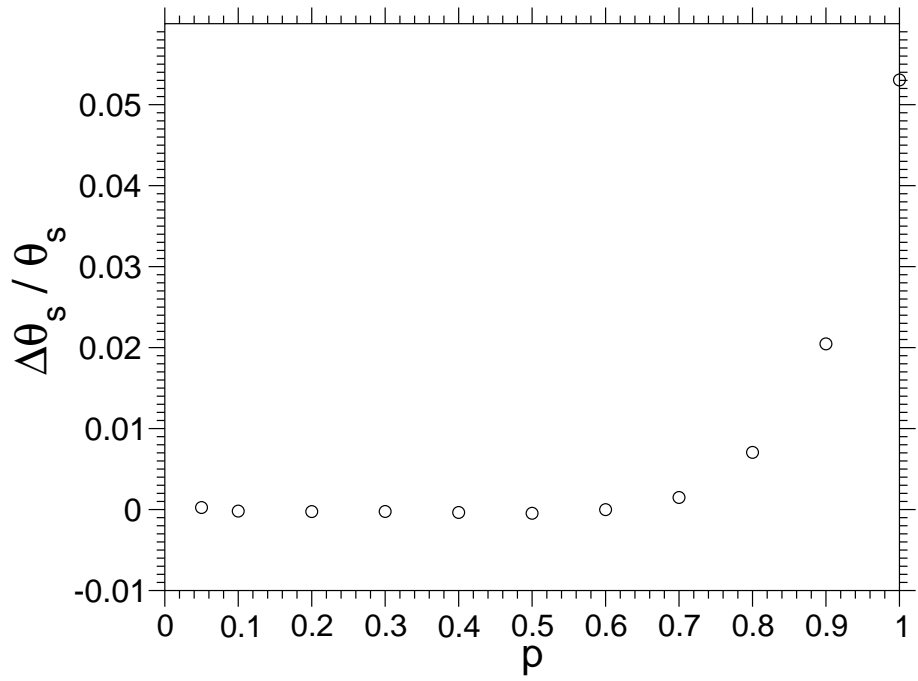}
\caption{\label{fig6} Relative error of $\theta_S$ as a function of 
$p$ in the second order approximation for the AR case. }
\end{figure}

An analytical solution for these equations does not seem possible, but
they can be integrated numerically. The results for the stationary coverage
are given in Tables \ref{TabCPD} and 
\ref{TabCTD}. They are significantly better for the CR case; the  
deviation from the numerical result is maximal for $p=1$ and is about 
$-0.4\%$; its absolute value $|\Delta \theta_S/\theta_S|$ diminishes 
monotonically with decreasing $p$. In contrast, the maximal deviation 
for $p=1$ in the AR case makes about $5.3\%$ (cf. Fig. \ref{fig6}). 

Better approximations can be obtained at higher orders, but the number
of conditional probabilities to be taken into account grows 
dramatically. It then becomes necessary to automate the generation of
the hierarchical equations. For instance, to third order one has 24 
different probabilities and to fourth order, 766 \cite{Nord}. 

\begin{figure}[ht]
\includegraphics[scale=0.6]{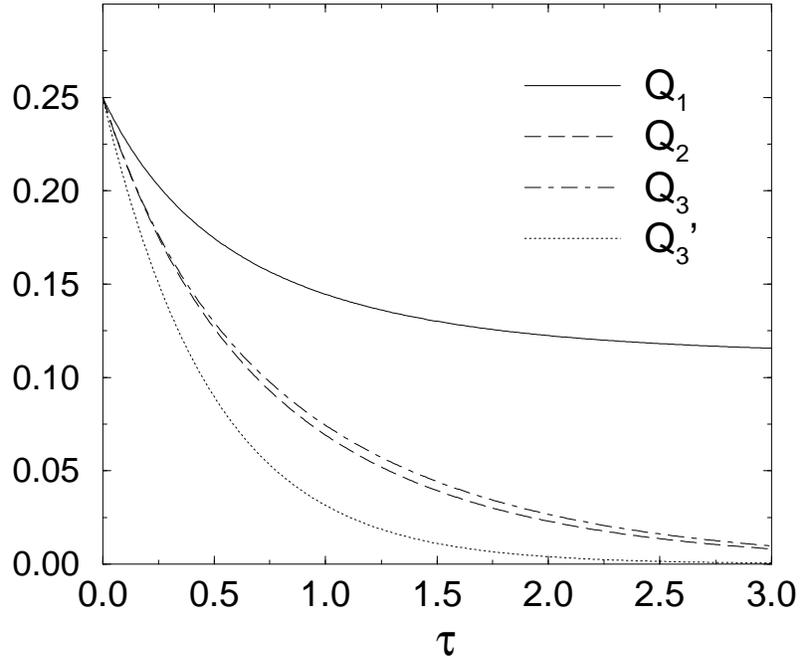}
\caption{\label{fig7} Time evolution of the conditional probabilities
obtained from Eqs. (\ref{secorhie})
for the AR case in the dilute regime ($p=0.25$).} 
\end{figure}

Nevertheless, the approximate conditional probabilities obtained 
from the second-order hierarchy (\ref{secorhie}) are already in good 
agreement with exact simulation results, both at the level of the 
stationary coverage and at the level of the time evolution (data not shown). 
Interestingly, the dynamics turns out to be qualitatively different 
depending on the value of $\nu$. In the CR case the inequality 
$Q_{\occs}>Q_{\inda}$ holds for all times, 
while in the AR case this is only true provided that 
the initial coverage is sufficiently low, i.e. for $p<1/2$. 
This behavior is observed in Fig. \ref{fig7}, which also displays
the time evolution of the other conditional probabilities 
(for typographical reasons, the symbols $Q_1, Q_2, Q_3$ and ${Q_3}^{'}$ used
in the legend represent respectively the quantities $Q_{\occs}, Q_{\inda}, 
Q_{\indc}$ and $Q_{\indk}$).  

\begin{figure}[ht]
\includegraphics[scale=0.6]{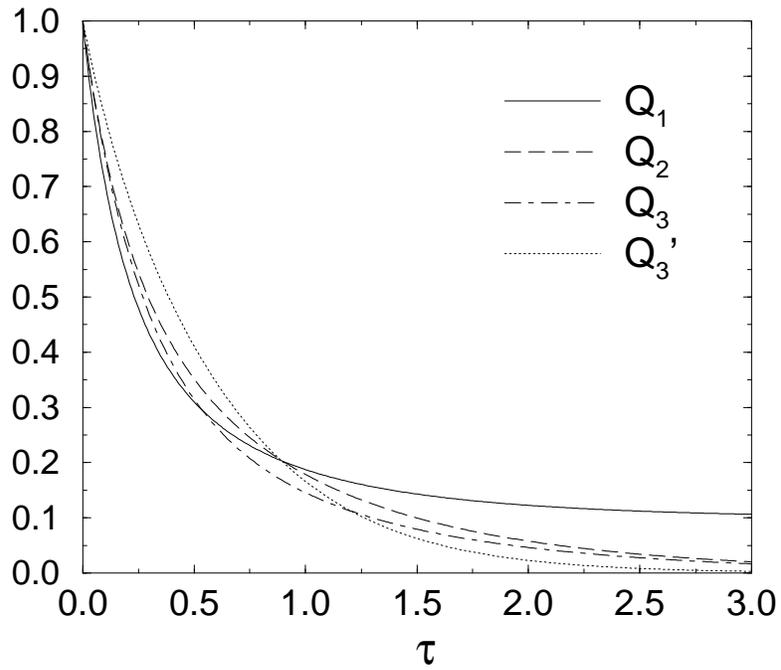}
\caption{\label{fig8} Time evolution of the conditional probabilities
obtained from Eqs. (\ref{secorhie})
for the AR case in the saturation regime ($p=1$). The legend uses the same
notation as in Fig. \ref{fig7}.} 
\end{figure}

In contrast, above $p=1/2$ the AR system displays a crossover between 
a short time regime for which 
$Q_{\occs}<Q_{\inda}$ and a long time regime with $Q_{\occs}>Q_{\inda}$
beyond a $p$-dependent crossover time (see Fig. \ref{fig8} ). 
I.e., for short times the probability to find a site occupied given that its 
neighbor is occupied is larger than for a randomly chosen site with no 
previous information on the state of the neighbor site, whereas for long 
times the opposite is true. Remarkably enough, the qualitative behavior 
of both reaction schemes appears to be universal, in the sense that it 
remains the same in Bethe lattices of arbitrary coordination number (cf. 
Eqs. \ref{condbeth}). 

As far as higher order conditional probabilities are concerned, 
the inequality $Q_{\occs} > Q_{\inda} > Q_{\indc} > Q_{\indk}$ holds 
at all times both in the CR case and in the dilute AR case with $p<1/2$ (cf. Fig.
\ref{fig7} ). 
However, for sufficiently short times we again observe a departure 
from this behavior at higher $p$ in the AR case. In fact, all three 
conditional probabilities  $Q_{\occs}, Q_{\inda}, Q_{\indc}$ 
become larger than $Q_{\occs}$  for a sufficiently large 
$p$ (cf. Fig. \ref{fig8} ). In this regime, the detailed behavior of the 
above $Q$ probabilities with respect to each other is rather complex 
and shall not be further discussed here.   

In order to interpret some of the above results,   
let us first characterize the occupation of a given site $i$ by an occupation
number $n_i$ (equal to one if the site is occupied and zero otherwise). The
fluctuation $\delta n_i$ is defined as the deviation from the average
occupation in a given statistical realization, i.e. 
$\delta n_i=n_i-\langle n_i \rangle$. A special kind of 
two-site fluctuation correlation is then 
$f_m \equiv \langle \, \delta n_i \, \delta n_{i+m} \rangle$,
where $i$ and $i+m$ are two sites separated by $m$ bonds along a $1D$ path. 
By definition, $f_m$ is translationally invariant and depends only on 
the distance $m$.  
 
\begin{figure}[ht]
\includegraphics[scale=0.7]{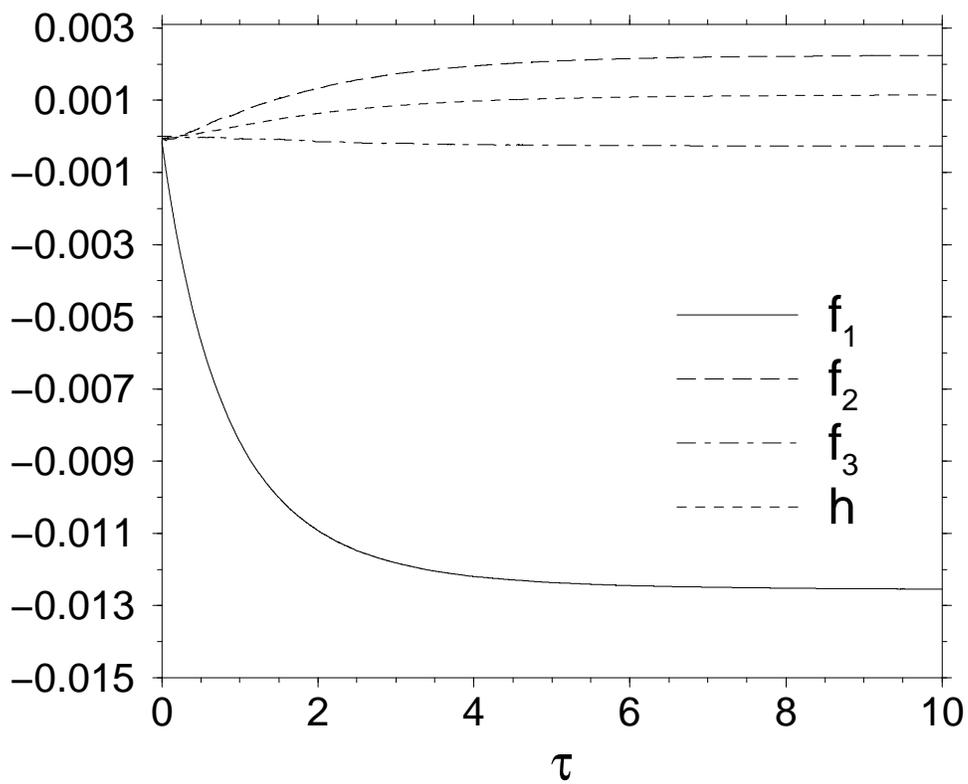}
\caption{\label{fig9} Dynamical behavior of the first three two-site 
fluctuation correlation functions and the three-site correlation
function $h$ in the dilute AR case ($p=0.25$). For this computation
we have performed simulations over 5000 realizations
on a $200\times 200$ lattice.}  
\end{figure}

\begin{figure}[ht]
\includegraphics[scale=0.7]{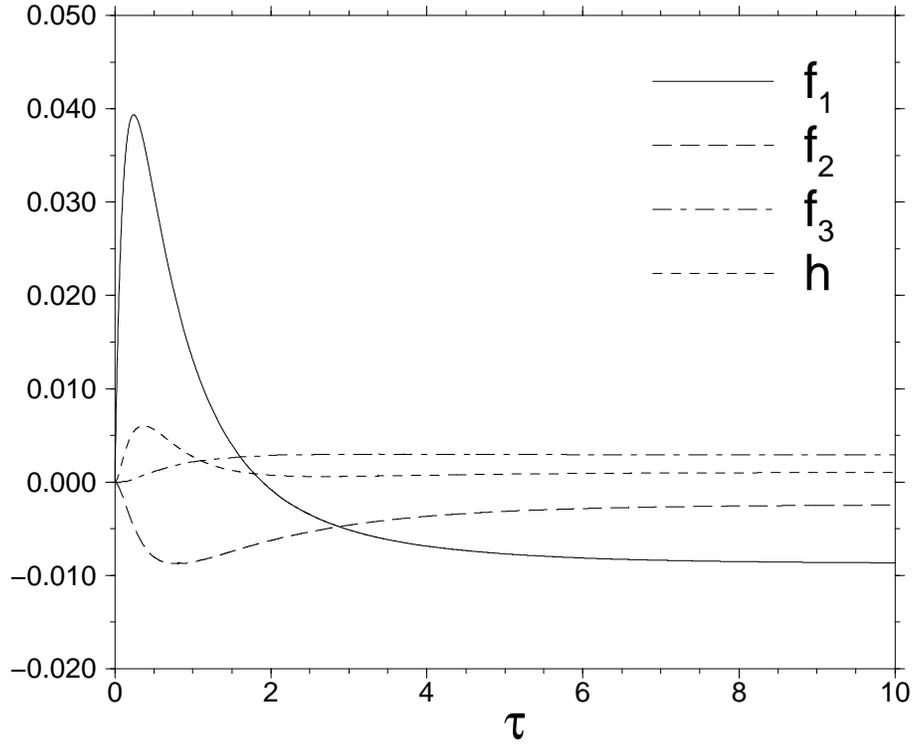}
\caption{\label{fig10} Dynamical behavior of the first three two-site 
fluctuation correlation functions and the three-site correlation
function $h$ in the AR case with an initially full lattice ($p=1$). 
The simulation parameters are the same as in Fig. \ref{fig9}.}
\end{figure}

The behavior of the conditional probabilities in our hierarchy is given by 
the cluster probabilities $P$. In turn, the latter   
are related to the fluctuation correlations, which measure the 
reaction-induced ordering in the system.
E.g., the sign of the difference $Q_{\occs}-Q_{\inda}$ is the same as 
the sign of the nearest neighbour two-site fluctuation correlation 
$f_1 = P_{\pina}-P_{\occs}^{\,\,2}=(Q_{\inda}-Q_{\occs})\,Q_{\occs}$. 
In all cases $f_1<0$ as $\tau\to\infty$, since two-particle clusters
disappear. In Fig. \ref{fig9} MC computations for  
the dynamical behavior of the two-site correlations 
$f_1,f_2, f_3$ and the three site correlation 
$h=\langle \delta n_i \, \delta  n_{i+1} \, \delta n_{i+2}\rangle$
in the dilute AR case ($p<\frac{1}{2}$) are shown. As in the CR case, one has
$f_1<0$ for all times, i.e $P_{\pina}> P_{\occs}^{\,\,2}$.
However, as soon as $p>1/2$, one has $f_1>0$
for sufficiently short times (see Fig \ref{fig10}). 
In other words, the probability to find a 
pair of neighboring sites simultaneously occupied is higher than 
if both sites are chosen at random. Most probably, the reason is that
for short times the typical size of particle islands is still
relatively large, and so is the value of $P_{\pina}$ ; however,
the reaction-induced growth of empty site clusters 
takes place at a higher rate than in the CR case.  
Thus, the probability that one finds an empty site beyond a certain 
correlation length from a given particle is comparatively high, thereby 
decreasing the value of $P_{\occs}^{\,\,2}$. 

As for the behavior of $f_2$,$f_3$ and $h$, both schemes again display 
very similar qualitative features in the low-$p$ regime. The numerical 
plots in Fig. \ref{fig9} suggest that 
$f_2 >0$ and $f_3<0$ for all times 
(for very short times, however, our precision does not allow to determine
the sign of the correlation functions). In any case, this 
holds for the stationary values of these quantities as $\tau\to \infty$. 
In terms of conditional 
probabilities, this means that $Q_{\indbuns}>Q_{\occs}$ and 
$Q_{\indhuns}< Q_{\occs}$, where ``-'' denotes a site in an unspecified 
state. Notice also that the absolute value of the 
three-site correlation $|h|$ becomes significantly 
larger than $|f_3|$. Moreover, for yet smaller values of the initial coverage
$|h|$ may get larger than $|f_2|$. This suggests that any expansion 
of the cluster probabilities retaining only two-site correlation 
functions fails to describe the behavior, 
since long-range correlations propagate throughout the system in the 
course of reaction. As a matter of fact, in the $1D$ case such an 
expansion leads to a zero stationary
coverage to any order of the distance between sites \cite{thes}.    
 
At higher values of $p$, the behavior is again modified in the AR case.
The functions $f_2$ and $f_3$ change sign, and the above 
inequalities for the conditional probabilities change their direction.
In contrast, $h$ keeps its positive sign but decreases strongly. 
 
The analysis presented in this subsection suggests that the nature 
of the spatial self-ordering as a function of the initial 
condition is rather complex (specially in the AR case)
and remains to be fully characterized. 

\section{Conclusions and outlook}

Using the analogy with RSA problems, we have used the method
of conditional probabilities to compute estimates for the lattice 
coverage in the framework of a unifying description for
two different types of irreversible binary reactions, i.e. coalescence
and annihilation. More traditional methods 
based on a spatial cut-off of fluctuations fail here, since the latter
are propagated by the reactions over the whole system size. 
In contrast, the method of conditional probabilities is exact
in $1D$ and branching media such as Bethe lattices, which can 
be used as a starting point for density expansions in other
regular lattices \cite{Evans3} (in the dilute limit, the Bethe 
lattice approximation should be good, since clusters with loops are
rare). A further advantage of the method is that it provides a 
reasonable approximation for the ``exact'' simulation results 
beyond the dilute limit, thereby allowing to obtain a fairly good
estimate in the vicinity of $p=1$ (corresponding to the usual
initial condition in RSA problems). Remarkably, the expansion
for the CR model in this regime provides a better approximation
than for the AR model. 

The approach used in the present paper can also be applied to mixed 
systems combining both coalescence and annihilation steps as well as
to more complex kinds of initial condition \cite{cadil}. The correspondence
between such models and RSA problems  
may prove useful in the context of pre-patterning 
of the substrate as a tool to improve self-assembly in certain systems. 

We have also seen that our model yields good results for the 
fluctuation-induced dynamical behavior of the system. The main conclusion 
is that the subsequent dynamics of the spatial distribution is very 
sensitive with respect to the details of the initial condition, specially
in the AR case, where several types of crossovers for the correlation 
functions have been identified. 

Possible extensions of our work include a more complete characterization
of the transient behavior of the spatial distribution for the
reactant species (and not only for the special kind of correlation
functions considered here) as well as its dependence on the initial
condition. However, exact decimation at any scale is in principle 
only possible in one dimension \cite{bonnier} and probably also 
on Bethe lattices, but the analogous problem on a lattice 
with loops still requires the use of approximate techniques. 

In the above context, it is also of interest to compare the properties
of such systems with those of their diffusion-controlled counterparts.
This work could then be further extended to other
systems such as the two-species annihilation $A+B\rightarrow S+S$.
This reaction is known to induce reactant segregation at low 
dimensions and has been widely investigated in the diffusion-controlled
case \cite{Ovch,Touss,Kuz,Bram,Schn,linden,sancho}, but its version
with immobile reactants \cite{Priv} has not received much attention yet.
In particular, it would be interesting to see whether
a shielding property can also be derived in this case, at least for a 
specific kind of initial conditions. 

\section*{ACKNOWLEDGMENTS}

I thank Prof. G. Nicolis for his careful reading of 
the manuscript as well as for helpful suggestions.

\newpage

\begin{table}[htbp]
\begin{ruledtabular}
\begin{tabular}{cccc}
$p$ & MC Simulation & truncation (1st. ord.) & truncation (2nd. ord.) \\[0.5cm]
\hline
0.05 &0.04535379   &0.04535147 &0.04535233\\
0.1  &0.08265606   &0.08264463 &0.08265592\\
0.2  &0.1390146    &0.1388889 &0.1390138\\
0.3  &0.1779778    &0.1775148 &0.1779583\\
0.4  &0.2050887    &0.2040816 &0.2050757\\
0.5  &0.2239603    &0.2222222 &0.2223642\\
0.6  &0.2369787    &0.2343750 &0.2369514\\
0.7  &0.2456631    &0.2422145 &0.2456311\\
0.8  &0.2510806    &0.2469136 &0.2510572\\
0.9  &0.2540033    &0.2493075 &0.2539584\\
1    &0.2549411    &0.2500000 &0.2548402\\
\end{tabular}
\end{ruledtabular}
\caption{\label{TabCPD} A comparison between the values of 
$\theta_S$ obtained from MC simulations and the
approximated values obtained by truncation of the hierarchy 
for the CR model.}
\end{table}

\begin{table}[htbp]
\begin{ruledtabular}
\begin{tabular}{cccc}
$p$ & MC Simulation & truncation (1st. ord.) & truncation (2nd. ord.) \\[0.5cm]
\hline
0.05  &0.04131765 &0.04132231  &0.04132796\\
0.1   &0.06952005  &0.06944444 &0.06950689\\
0.2   &0.1025630   &0.1020408  &0.1025378\\
0.3   &0.1185036   &0.1171875  &0.1184757\\
0.4   &0.1255734   &0.1234568  &0.1255286\\
0.5   &0.1274785   &0.1250000  &0.1274201\\
0.6   &0.1259282   &0.1239669  &0.1259275\\
0.7   &0.1217432   &0.1215278  &0.1219250\\
0.8   &0.1150304   &0.1183432  &0.1158427\\
0.9   &0.1057193   &0.1147959  &0.1078826\\
1     &0.09318323  &0.1111111  &0.09812664\\
\end{tabular}
\end{ruledtabular}
\caption{\label{TabCTD} A comparison between the values of 
$\theta_S$ obtained from MC simulations and the
approximated values obtained by truncation of the hierarchy 
for the AR model.}
\end{table}

\end{document}